\documentclass[twocolumn]{aastex6}
\usepackage{graphicx}
\usepackage{amssymb,amsfonts,amsmath,amstext,amsgen,amsopn,amsxtra,indentfirst,times,longtable}

\begin{document}

\title{The Transient Outgassed Atmosphere of 55 Cancri e}

\author{Kevin Heng\altaffilmark{1,2,3}}
\altaffiltext{1}{Faculty of Physics, Ludwig Maximilian University, Scheinerstrasse 1, D-81679, Munich, Bavaria, Germany. Email: Kevin.Heng@physik.lmu.de}
\altaffiltext{2}{ARTORG Center for Biomedical Engineering Research, University of Bern, Murtenstrasse 50, CH-3008, Bern, Switzerland}
\altaffiltext{3}{Astronomy \& Astrophysics Group, Department of Physics, University of Warwick, Coventry CV4 7AL, United Kingdom}

\begin{abstract}
The enigmatic nature of 55 Cancri e has defied theoretical explanation.  Any explanation needs to account for the observed variability of its secondary eclipse depth, which is at times consistent with zero in the visible/optical range of wavelengths---a phenomenon that does not occur with its also variable infrared eclipses.  Yet, despite this variability its transit depth remains somewhat constant in time and is inconsistent with opaque material filling its Hill sphere.  The current study explores the possibility of a thin, transient, secondary atmosphere on 55 Cancri e that is sourced by geochemical outgassing.  Its transient nature derives from the inability of outgassing to be balanced by atmospheric escape.  As the outgassed atmosphere escapes and is replenished, it rapidly adjusts to radiative equilibrium and the temperature fluctuations cause the infrared eclipse depths to vary.  Atmospheres of pure carbon dioxide or carbon monoxide produce sufficient Rayleigh scattering to explain the observed optical/visible eclipse depths, which vanish in the absence of an atmosphere and the presence of a dark rocky surface.  Atmospheres of pure methane are ruled out because they produce insufficient Rayleigh scattering.  Upcoming observations by the James Webb Space Telescope will potentially allow the atmospheric temperature and surface pressure, as well as the surface temperature, to be measured.
\end{abstract}


\section{Introduction}
\label{sect:intro}

Since the discovery of its transits in 2011 \citep{demory11}, the super Earth 55 Cancri e is one of the most enigmatic objects in the study of exoplanetary atmospheres.  Searches for hydrogen \citep{ehrenreich12} and helium \citep{zhang21} in its atmosphere have been unsuccessful, which is not inconsistent with its bulk mass density of $6.4^{+0.8}_{-0.7}$ g cm$^{-3}$ \citep{demory16b}.  Models of its interior structure suggest the possibility of volatiles being present \citep{dorn17,crida18}, but to date an unambiguous detection of gaseous species in its atmosphere remains elusive \citep{jindal20} including a search for metals using high-resolution spectroscopy from the ground \citep{keles23}.  A detection of hydrogen cyanide (HCN) was previously claimed \citep{tsiaras16}, but from a chemical perspective it remains difficult to reconcile with the non-detection of hydrogen \citep{ehrenreich12}.  

A Spitzer Space Telescope phase curve of 55 Cancri reveals a strong dayside ($\approx 2700$ K) to nightside ($\approx 1400$ K) brightness temperature contrast at 4.5 $\mu$m and a $41 \pm 12$ degree offset in the peak of its phase curve \citep{demory16b}.  (See also \citealt{mercier22}.)  This Spitzer phase curve has inspired several theoretical studies that have attempted to explain and/or predict its atmospheric composition \citep{ah17,hp17,maha17,miguel19}, although none of the predicted atmospheric chemistries have been definitively corroborated by the observations.

\begin{figure*}[!ht]
\begin{center}
\vspace{-0.1in} 
\includegraphics[width=\columnwidth]{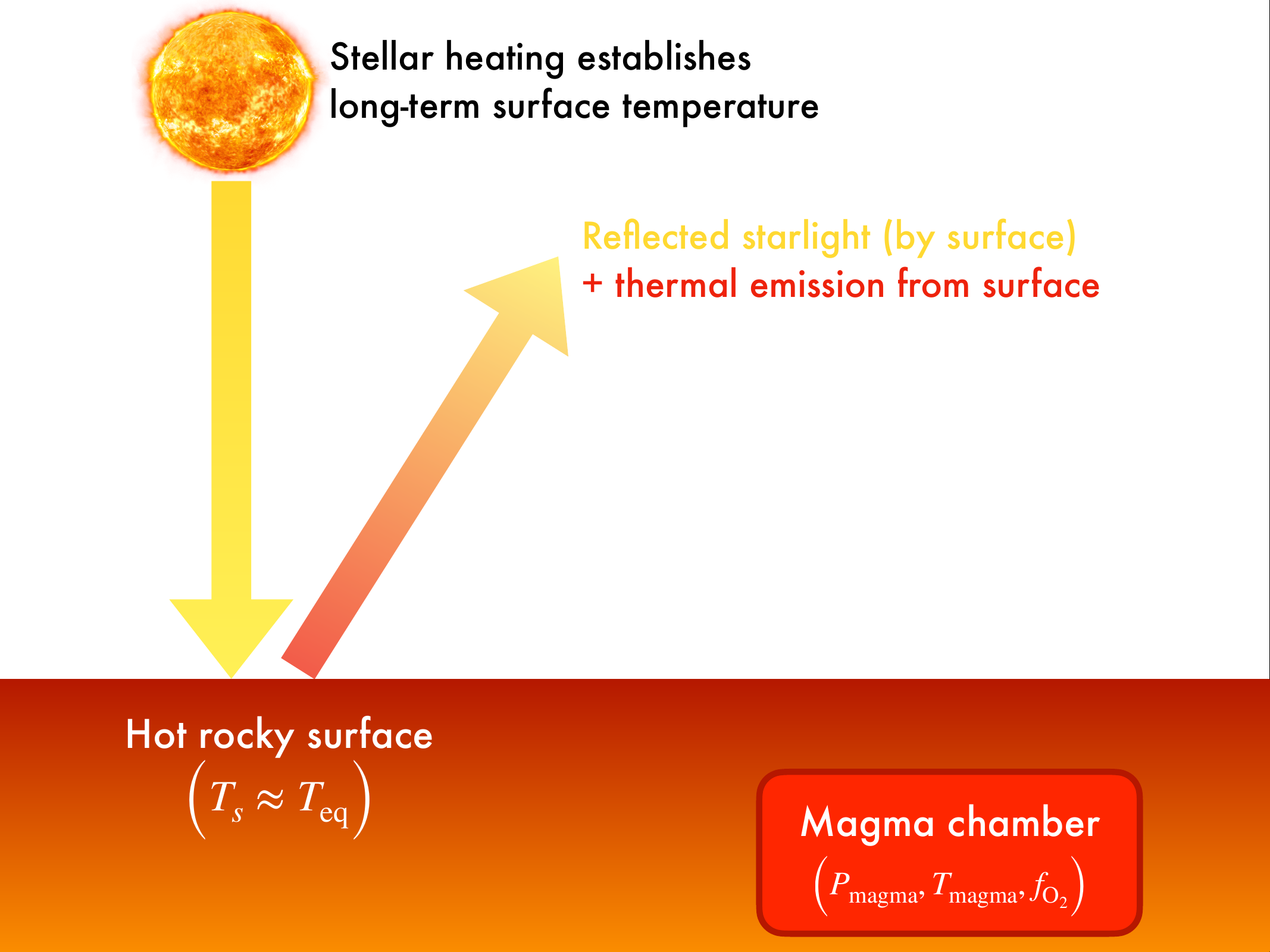}
\includegraphics[width=\columnwidth]{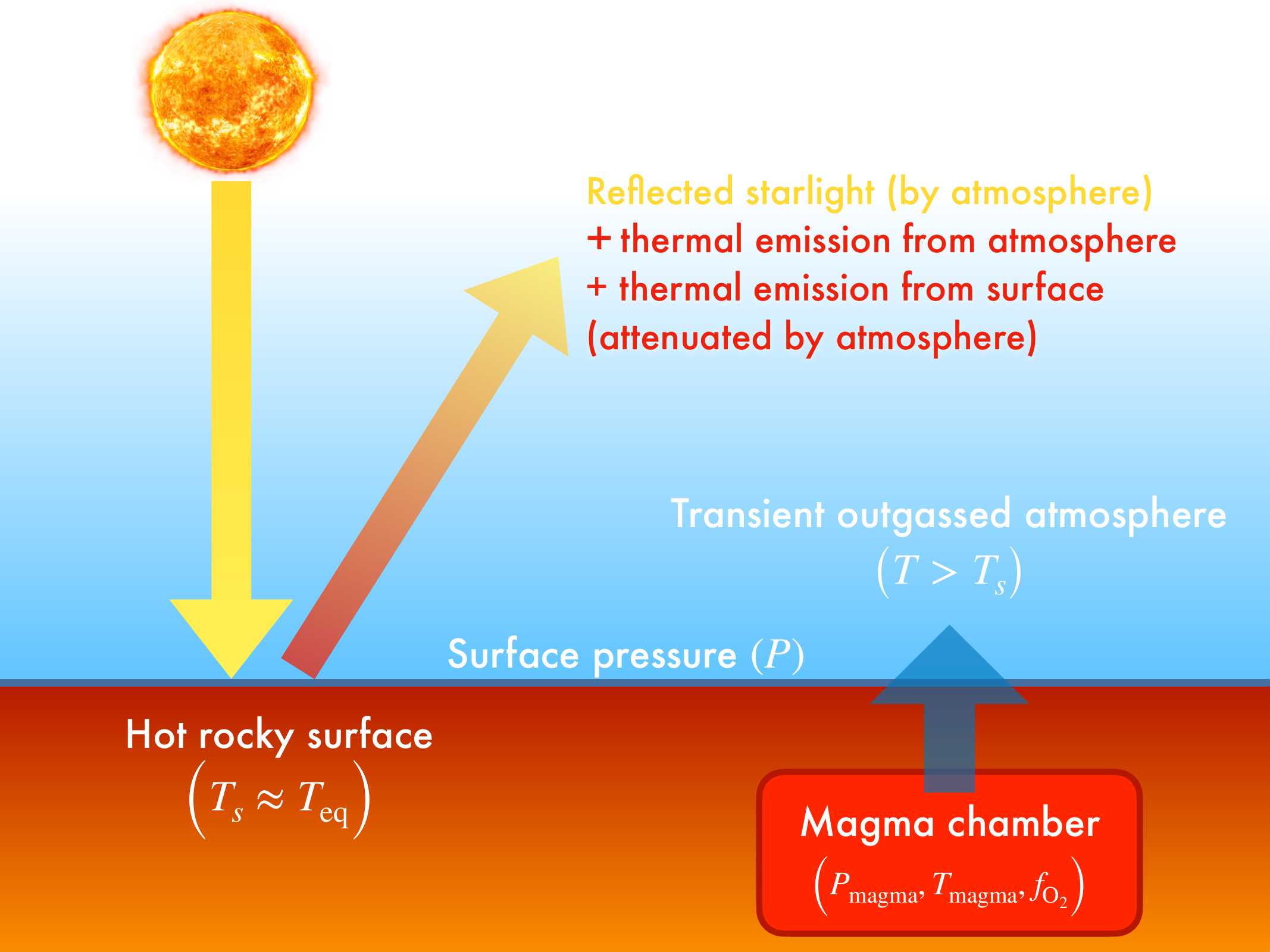}
\end{center}
\caption{Schematic of multi-wavelength model that includes a hot rocky surface (heated by starlight) and a thin, transient atmosphere sourced by geochemical outgassing.  The left and right panels show the model in its ``bare rock" and thin-atmosphere phases, respectively.  The atmosphere is transient because outgassing and atmospheric escape fail to balance out each other and the timescale for adjustment to radiative equilibrium is likely to be shorter than an orbital period (see text for discussion).}
\label{fig:schematic}
\end{figure*}

A key hint to understanding the nature of 55 Cancri e lies in the variability of its secondary eclipses, both in the optical/visible \citep{mv22,mv23,demory23} and infrared \citep{demory16a,tamburo18} range of wavelengths.  At times, the optical/visible secondary eclipses become consistent with zero \citep{demory23,mv23}, but this phenomenon does not occur with the infrared secondary eclipses \citep{demory16a,tamburo18}.  Yet, its primary transit depths remain somewhat constant \citep{tamburo18,mv23}.  Star-exoplanet interactions have been ruled out \citep{morris21}.  By analogy with the torus around Io (a moon of Jupiter), it has been suggested that a torus of spectroscopically active gas and dust could account for the observed variability of the secondary eclipses of 55 Cancri e \citep{mv23}.  

In addition to the generally short sublimation timescales of dust \citep{mv23}, a key puzzle of this interpretation is how the presence of a torus could be simultaneously consistent with the somewhat constant transit depths and variable eclipse depths.  Using the exoplanetary properties from \cite{demory16b} and the stellar properties from \cite{crida18}, the Hill radius of 55 Cancri e is estimated to be about $7.4 R_\oplus$ or about $3.9 R$ where $R=1.91 R_\oplus$ is the Spitzer $4.5 ~\mu$m radius of the exoplanet \citep{demory16b}.  This radius corresponds to a transit depth of about 334 parts per million (ppm), which is consistent with the 29 transit depths measured by the CHEOPS space telescope \citep{mv23}.  By contrast, the presence of a torus around 55 Cancri e, which would fill its Hill sphere, would result in a transit depth of about 4995 ppm, which is firmly ruled out by the CHEOPS observations.  It is worth noting that the Hill radius of Io is about 6 times its radius and its torus has been observed to fill and even overflow its Hill sphere (e.g., \citealt{st95,steffl04}).

One way out of this conundrum is to assume that spectroscopically active material is present only near secondary eclipse (where the dayside of the exoplanet is in full view), but somehow vanishes at primary transit (which probes its nightside and teminators).  Yet another way out is to assume that the material is spectroscopically active only at secondary eclipse, but becomes inert or transparent at primary transit.  From the perspective of Occam's Razor, it is challenging to explain how this may occur over the 29 contemporaneous transits and eclipses of 55 Cancri e observed by CHEOPS \citep{mv23}.

These recently reported observations by the CHEOPS space telescope \citep{mv22,mv23,demory23}, as well as upcoming eclipse observations by the James Webb Space Telescope (JWST)\footnote{JWST Cycle 1 Proposals \#1952 and \#2084.}, motivate a qualitatively different approach of interpreting the data of 55 Cancri e.  In the current study, it is suggested that the reported observations collectively describe a thin, transient, outgassed atmosphere that does not survive over long timescales because atmospheric escape dominates the outgassing flux.

\section{Toy Model of Transient Secondary Atmosphere}
\label{sect:method}

\subsection{Theory}

\begin{figure*}
\begin{center}
\vspace{-0.2in}
\includegraphics[width=\columnwidth]{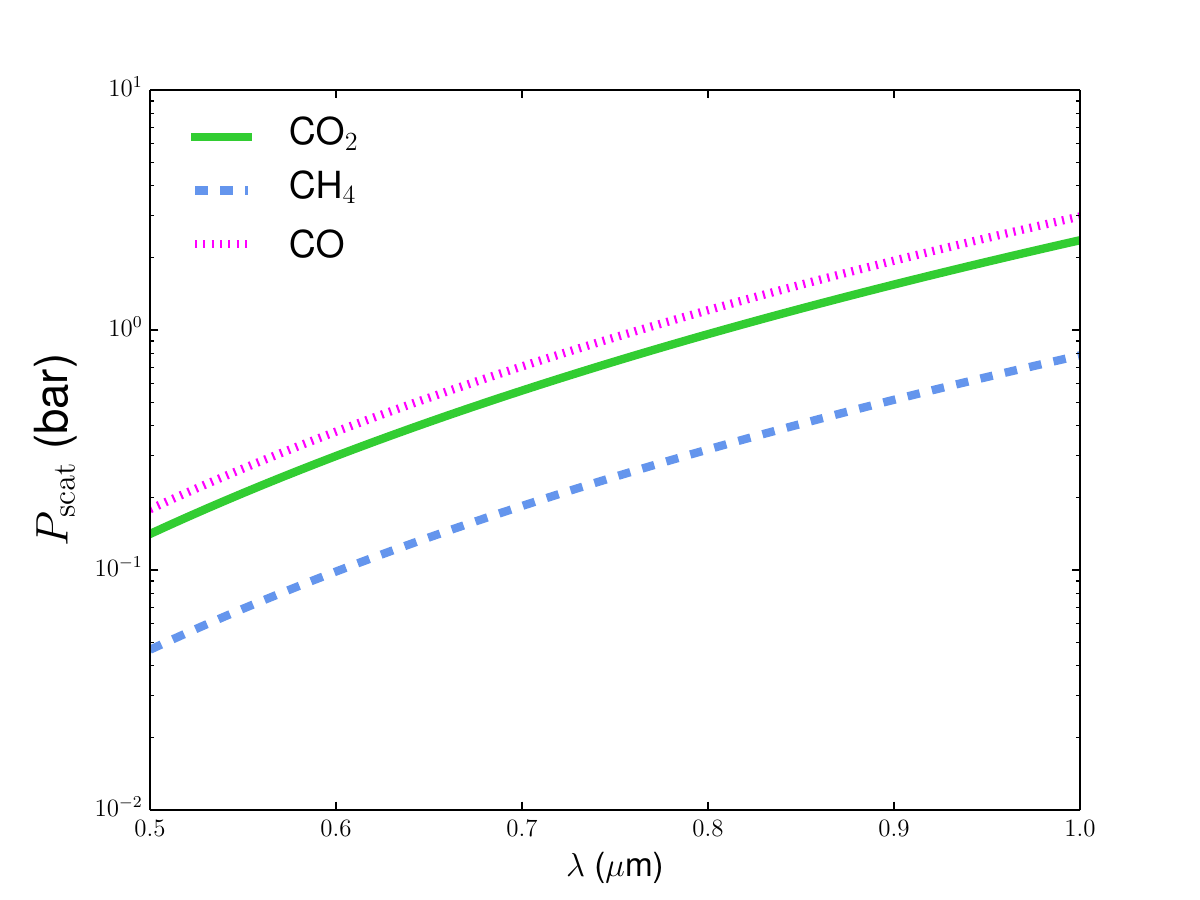}
\includegraphics[width=\columnwidth]{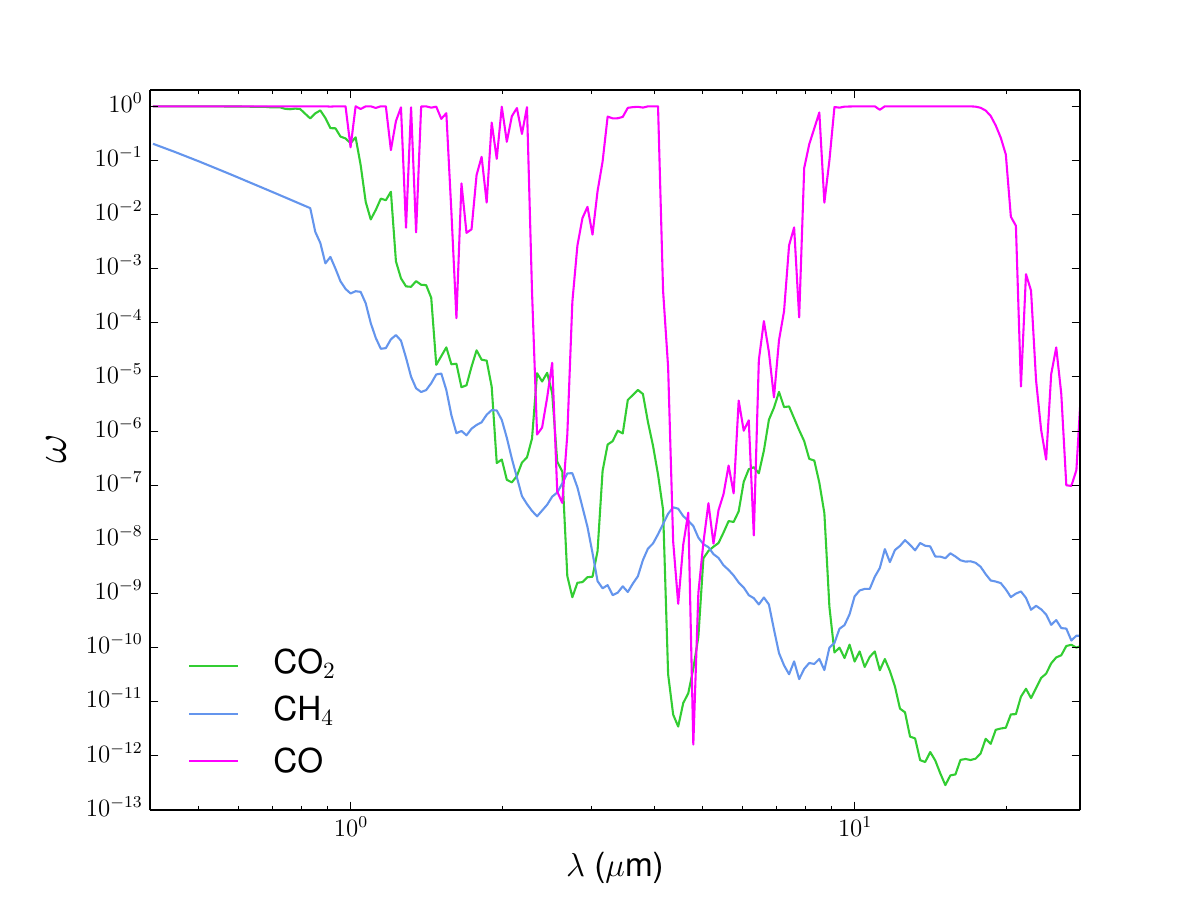}
\includegraphics[width=\columnwidth]{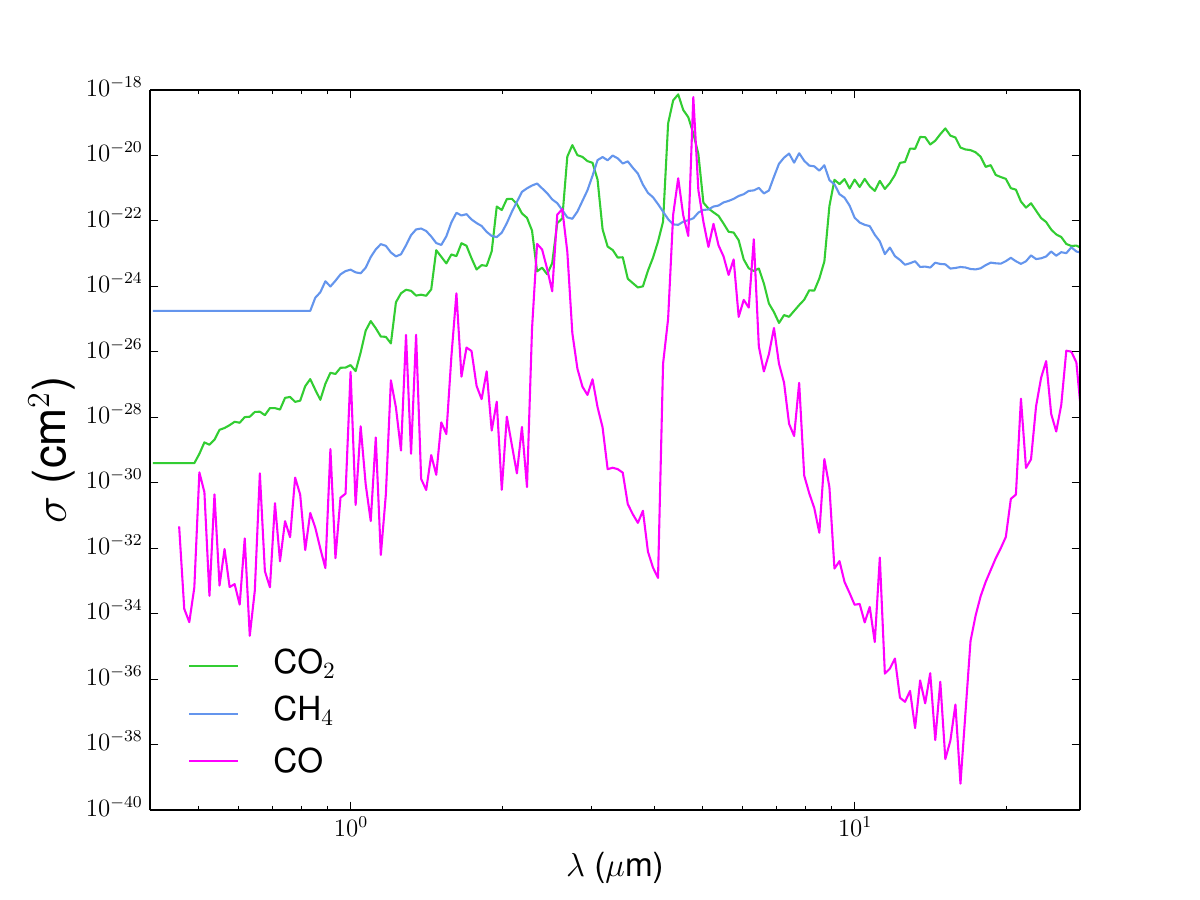}
\includegraphics[width=\columnwidth]{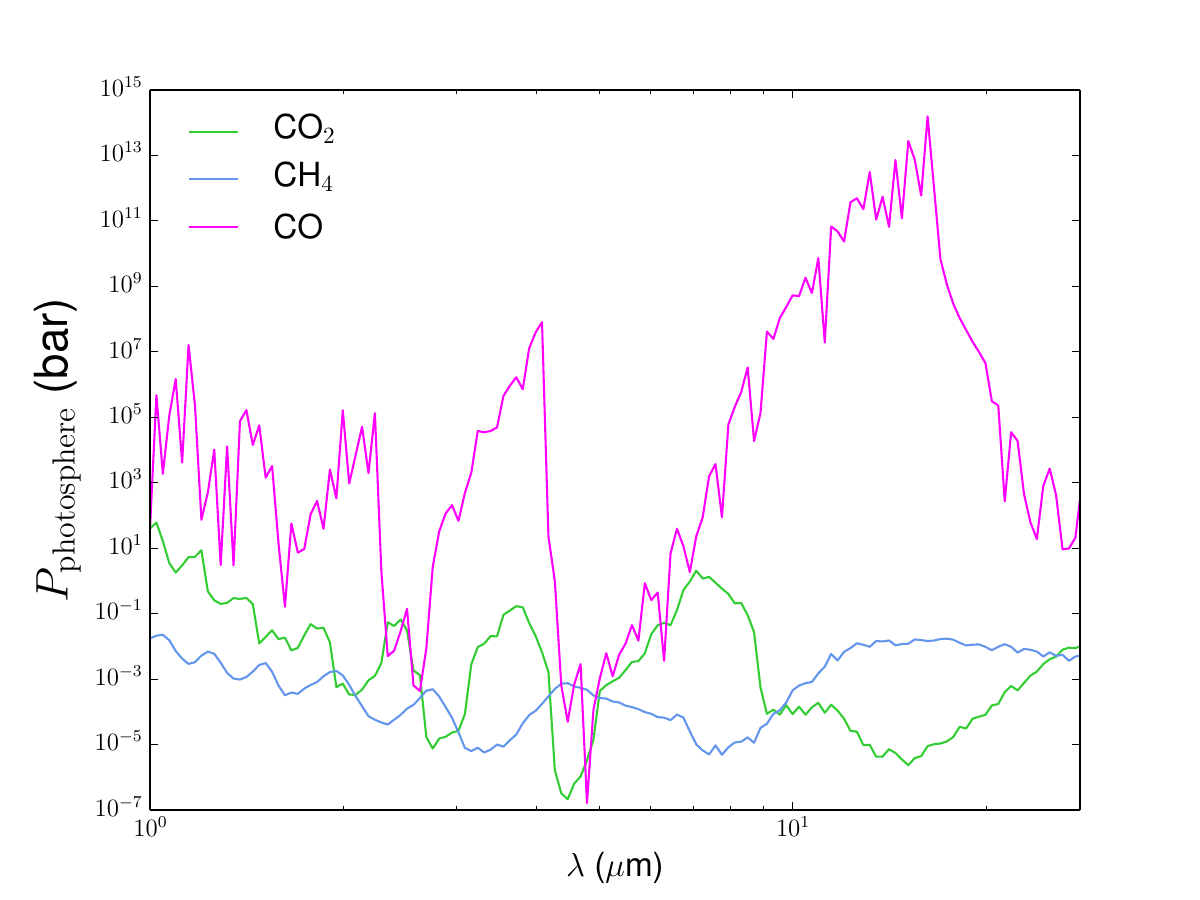}
\end{center}
\vspace{-0.1in}
\caption{Scattering photospheric pressures (top left panel), single-scattering albedos (top right panel), absorption cross sections (bottom left panel) and absorption photospheric pressures (bottom right panel) associated with pure carbon dioxide (CO$_2$), carbon monoxide (CO) and methane (CH$_4$) atmospheres.  For the absorption cross sections, a temperature of 2700 K and a pressure of 0.1 bar are assumed.}
\label{fig:quantities}
\end{figure*}

As a proof of concept, I will construct the simplest possible model that includes the salient features needed to explain multi-wavelength observations (Figure \ref{fig:schematic}).  The simplicity of the model is also motivated by the complexity of the processes that one needs to model.  The starting point is a secondary atmosphere sourced by geochemical outgassing (e.g., \citealt{th23}).  While the outgassing chemistry may be solved by considering the thermodynamics of mixed phases and non-ideal gases \citep{french66,gs14,th23}, predicting the outgassing \textit{flux} is challenging as it requires an understanding of the interior geodynamics of the exoplanet including its tectonic regime \citep{ss16}.  The outgassing is balanced by a process that is notoriously difficult to model even for the planets of the Solar System: atmospheric escape \citep{sa96}.  It is the balance between the outgassing and escape fluxes that allows one to calculate the atmospheric surface pressure (e.g., \citealt{liggins20}).  Instead of introducing several free parameters into the problem, I \textit{choose} to encode our incomplete knowledge of these processes into a single free parameter: the atmospheric surface pressure $P$.  

Furthermore, it is assumed that the atmospheric escape flux exceeds the outgassing flux such that an equilibrium between the two processes is never reached.  Such an assumption is consistent with the erratic temporal variability of the observed secondary eclipses, which is not observed to correlate with any exoplanetary property \citep{mv23}.  Using dimensional analysis, the energy-limited mass flux\footnote{Technically, this is the ``mass luminosity".} of atmospheric escape is
\begin{equation}
\dot{M} \sim \frac{\pi R^2 F_X}{E_g} = \frac{L_X R^3}{4 GM R^2_\star} \sim 10^{10} \mbox{ g s}^{-1},
\end{equation}
where $F_X$ is the X-ray flux of the star, $L_X = 4 \times 10^{26}$ erg s$^{-1}$ is its X-ray luminosity \citep{ehrenreich12}, $E_g=GM/R$ is the specific gravitational potential energy, $G$ is the universal gravitational constant, $M=8.08 ~M_\oplus$ is the mass of 55 Cancri e \citep{demory16b} and $R_\star = 0.958 ~R_\odot$ is the stellar radius \citep{crida18}.  This escape flux is $\sim 1000$ times higher than the CO$_2$ outgassing flux of modern Earth \citep{pm19}.  While a predictive theory for estimating the outgassing flux on 55 Cancri e remains elusive \citep{meier23}, it seems implausible that such a high outgassing flux could be attained.  To balance atmospheric escape, the mass of atmosphere outgassed in an orbital period ($t_{\rm orbit} = 0.74$ day) is $\dot{M} t_{\rm period} \approx 8 \times 10^{14}$ g.  This implies an atmospheric pressure of $\dot{M} t_{\rm period} g/ 4 \pi R^2 \approx 0.1 ~\mu$bar, which is orders of magnitude lower than the surface pressures assumed in this study.  The only general conclusion one can draw from the non-balancing of the two processes is that it produces stochasticity in the outgassing and atmospheric escape fluxes, which plausibly lead to temperature fluctuations.

The transient, outgassed atmosphere sits above a rocky surface, which is heated by starlight from 55 Cancri.  The equilibrium temperature of the exoplanet is about 1965 K.  In the absence of an atmosphere, the rocky surface attains a temperature of $T_s=1965$ K---at least, on the dayside of 55 Cancri e.  The thermal conduction timescale\footnote{While the thermal diffusivity of rock (which has the physical units of a diffusion coefficient) is well-documented, one needs to specify a characteristic length scale to convert it into a timescale.  It is unclear what the characteristic length scale is for a rocky surface.} associated with the rocky surface is assumed to be much longer than the survival time of the atmosphere, such that it does not have enough time to adjust to the atmospheric temperature.  The surface is assumed to radiate like a blackbody.  

The outgassed atmosphere originates from a magma chamber beneath the surface that is associated with its own temperature $T_{\rm magma}$.  Under Earth-like conditions, $T_{\rm magma} = 1600$ K; see \cite{th23} for a discussion of the appropriate values to assume for $T_{\rm magma}$.  Initially, the outgassed atmosphere has $T \approx T_{\rm magma}$.  Given enough time, the atmosphere adjusts its temperature to a value that is consistent with the greenhouse warming effect of the outgassed species.  This radiative adjustment time is \citep{sg02}
\begin{equation}
t_{\rm rad} \sim \frac{c_P P}{\sigma_{\rm SB} g T^3} \approx 2 \times 10^3 \mbox{ s} ~\left( \frac{P}{0.1 \mbox{ bar}} \right) ~\left( \frac{T}{1600 \mbox{ K}} \right)^{-3},
\end{equation}
where $c_P \sim 10^7$ erg g$^{-1}$ K$^{-1}$ is the specific heat capacity at constant pressure and $\sigma_{\rm SB}$ is the Stefan-Boltzmann constant.  The preceding estimate was made for CO$_2$ molecules and is a factor $\sim 2$--3 higher for CO and CH$_4$; the composition dependence enters through the specific heat capacity.  If the atmospheric surface pressure is $\sim 0.1$ bar, then $t_{\rm rad}$ is a fraction of the $0.74 \mbox{ day} \approx 6 \times 10^4$ s orbital period of 55 Cancri e.  It implies that, as soon as the atmosphere is outgassed, the adjustment of the atmospheric temperature to radiative equilibrium occurs well within an orbital period.

The key parameter controlling the chemistry of the outgassed atmosphere is the pressure associated with the magma $P_{\rm magma}$.  If the magma chamber is located close to the surface, then one may assume $P_{\rm magma} \approx P$ \citep{gs14}.  If it is located deep beneath the surface such that $P_{\rm magma} \gg P$, then the system inherits an additional free parameter.  An additional unknown, which is difficult to calculate from first principles, is the oxidation state of the mantle.  It is parametrised by the oxygen fugacity $f_{\rm O_2}$ (e.g., \citealt{th23}).  In the conceivable parameter space of $P_{\rm magma}$, $T_{\rm magma}$ and $f_{\rm O_2}$, the plausible carbon- and oxygen-bearing species are carbon dioxide (CO$_2$), carbon monoxide (CO), methane (CH$_4$) and water (H$_2$O) (e.g., \citealt{gs14,th23}).  In particular, a methane-dominated atmosphere is only produced for reduced (poorly oxidized) mantles, reduced magma temperatures (compared to that of Earth) and $P_{\rm magma} \gtrsim 10$ bar \citep{th23}.  Water is not considered further, because it is neither detected by the Wide Field Camera 3 of the Hubble Space Telescope \citep{tsiaras16} nor from the ground using high-resolution spectroscopy \citep{jindal20}.

Let the flux from the dayside of 55 Cancri e and the star be $F$ and $F_\star$, respectively.  The orbital semi-major axis of 55 Cancri e is $a=0.01544$ AU and the ratio of its radius to the stellar radius is $R/R_\star = 0.0187$ \citep{demory16b}.  It follows that the monochromatic secondary eclipse depth is
\begin{equation}
\frac{F}{F_\star} = \left( \frac{R}{R_\star} \right)^2 \frac{B_\lambda \left(T_s\right) {\cal T} + \left( 1 - {\cal T} \right) B_\lambda\left(T\right)}{B_\lambda \left(T_\star \right)} + \left( \frac{a}{R} \right)^2 A_g,
\label{eq:master}
\end{equation}
where $B_\lambda$ is the Planck function, $A_g$ is the geometric albedo and
\begin{equation}
{\cal T} = e^{-\tau}
\end{equation}
is the transmission function.  The effective temperature of the star is $T_\star = 5174$ K \citep{crida18}.

In equation (\ref{eq:master}), the first term accounts for thermal emission from the rocky surface, which is attenuated by the transient outgassed atmosphere, as well as from the atmosphere itself.  The term associated with $B_\lambda(T)$ is a well-known expression used to estimate the brightness temperature (e.g., \citealt{ca11,tamburo18}).  In the limit of an opaque atmosphere (${\cal T}=0$), the thermal emission is purely blackbody in nature (with no spectral features).  The second term in equation (\ref{eq:master}) accounts for reflected light from the outgassed atmosphere.  The optical depth associated with the attenuation is
\begin{equation}
\tau = \frac{\kappa P}{g},
\end{equation}
where $g=10^{3.33} \approx 2138$ cm s$^{-2}$ is the surface gravity of 55 Cancri e \citep{demory16b}.  The absorption opacity $\kappa$ is trivially converted to a cross section $\sigma$ via $\kappa = \sigma/m$, where $m$ is the mass of the molecular species considered.  

For Rayleigh scattering by molecules, the geometric albedo is given by \citep{heng21}
\begin{equation}
A_g = \frac{\omega}{16} + \frac{\epsilon}{2} + \frac{\epsilon^2}{6} + \frac{\epsilon^3}{24}, 
\label{eq:Ag}
\end{equation}
where $\epsilon = (1-\gamma)/(1+\gamma)$ is the bihemispherical reflectance \citep{hapke81} and $\gamma = \sqrt{1-\omega}$.  The single-scattering albedo $\omega$ is constructed from the absorption ($\sigma$) and scattering ($\sigma_{\rm scat}$) cross sections,
\begin{equation}
\omega = \frac{\sigma_{\rm scat}}{\sigma + \sigma_{\rm scat}}.
\end{equation}
If only single scattering is considered (i.e., multiple scattering is ignored), then the geometric albedo is instead given by \citep{heng21}
\begin{equation}
A_g = \frac{3\omega}{16}.
\label{eq:Ag_single}
\end{equation}

A more accurate approach is to derive an analytical expression for the geometric albedo that is a function of the scattering optical depth, which would allow an entire continuum of surface pressures to be rigorously considered.  The geometric albedo should naturally vanish as the scattering optical depth becomes zero.  While this development is necessary for future work, it is beyond the scope of the current study.

In summary, the toy model described here has only three free parameters: the surface temperature $T_s$, the atmospheric temperature $T$ and the atmospheric surface pressure $P$.  Each of the aforementioned chemical species will be considered in turn.  By fitting this model to data using Bayesian inference methods, one may derive posterior distributions of these three parameters.  This is beyond the scope of the current study.  Instead, the intention is to elucidate the influence of each parameter in an intuitive way.


\subsection{Cross sections}

Indispensable ingredients for even a toy model are the cross sections associated with the absorption and scattering of radiation.  The Rayleigh scattering cross sections for CO, CO$_2$ and CH$_4$ are taken from \cite{su05} and \cite{thalman14}.  Unfortunately, a literature search for the Rayleigh scattering of sulfur dioxide (SO$_2$) was in vain.  The spectroscopic line lists and partition functions are taken from the following references: \cite{li15} for CO; \cite{y20} for CO$_2$; \cite{yt14} and \cite{y17} for CH$_4$.  These quantities were converted into opacities (cross sections per unit mass) using the \texttt{HELIOS-K} opacity calculator \citep{gh15} and made publicly available via the DACE database \citep{grimm21}.\footnote{\texttt{https://dace.unige.ch/}}

\section{Results}
\label{sect:results}

In a self-consistent radiative transfer model, equation (\ref{eq:master}) is iterated with the first law of thermodynamics (energy conservation) such that radiative equilibrium is attained.  The outcome of such an approach is a temperature-pressure profile that is consistent with the assumed chemical abundances and emergent spectrum.  In atmospheric retrieval models of exoplanets, this approach is not typically implemented, i.e., the chemical abundances and temperature-pressure profile are not self-consistent.  This is the phenomenological approach that we will adopt when computing our toy model.  As already explained, the intention is to demonstrate the influence of each parameter rather than fit for its values from matching data.  As an illustration, the atmospheric temperature is assumed to be $T=2700$ K, which is inspired by the Spitzer 4.5 $\mu$m brightness temperature reported by \cite{demory16b}.  The elucidated concepts are qualitatively independent of this assumption.

\subsection{Scattering and absorption photospheres}

The expression for the geometric albedo stated in equation (\ref{eq:Ag}) assumes a so-called ``semi-infinite atmosphere", where the scattering optical depth spans the range from zero to infinity \citep{chandra,hapke81}.  Physically, this corresponds to a scattering atmosphere that transitions from being transparent to opaque, rather than the spatial distance going to infinity in one direction.  To check this assumption, I examine the scattering photospheric pressure,
\begin{equation}
P_{\rm scat} \sim \frac{g}{\kappa_{\rm scat}},
\end{equation}
where $\kappa_{\rm scat}$ is the scattering opacity.  The top left panel of Figure \ref{fig:quantities} shows calculations of $P_{\rm scat}$ for various assumed atmospheric compositions.  At wavelengths relevant to the spectral energy distribution of 55 Cancri ($\gtrsim 0.55 ~\mu$m), the scattering optical depth becomes unity at $\sim 0.1$ bar.  This suggests that as long as atmospheric surface pressures $\gtrsim 0.1$ bar are considered then the use of equation (\ref{eq:Ag}) is not unreasonable.  Otherwise, one ignores multiple scattering as an approximation and uses equation (\ref{eq:Ag_single}) instead.

The top right panel of Figure \ref{fig:quantities} shows the single-scattering albedos.  The absorption cross sections are shown in the bottom left panel of Figure \ref{fig:quantities}.  The (absorption) photospheric pressure,
\begin{equation}
P_{\rm photosphere} \sim \frac{g}{\kappa},
\end{equation}
is shown in the bottom right panel of Figure \ref{fig:quantities}.  The comparatively transparent spectral windows in between the absorption bands of CO means that the single-scattering albedo easily reaches unity within these windows.  It also means that $P_{\rm photosphere} \gg 1$ bar within the same spectral windows, implying that starlight may easily reach the surface of an exoplanet with a $\sim 0.1$ bar CO atmosphere.

By contrast, methane is a good absorber in the infrared range of wavelengths and does not have deep spectral windows in between its absorption bands.  An atmosphere with even $\sim 1$ mbar of methane may easily absorb infrared radiation from the surface of 55 Cancri e.

\begin{figure}
\vspace{-0.1in}
\begin{center}
\includegraphics[width=\columnwidth]{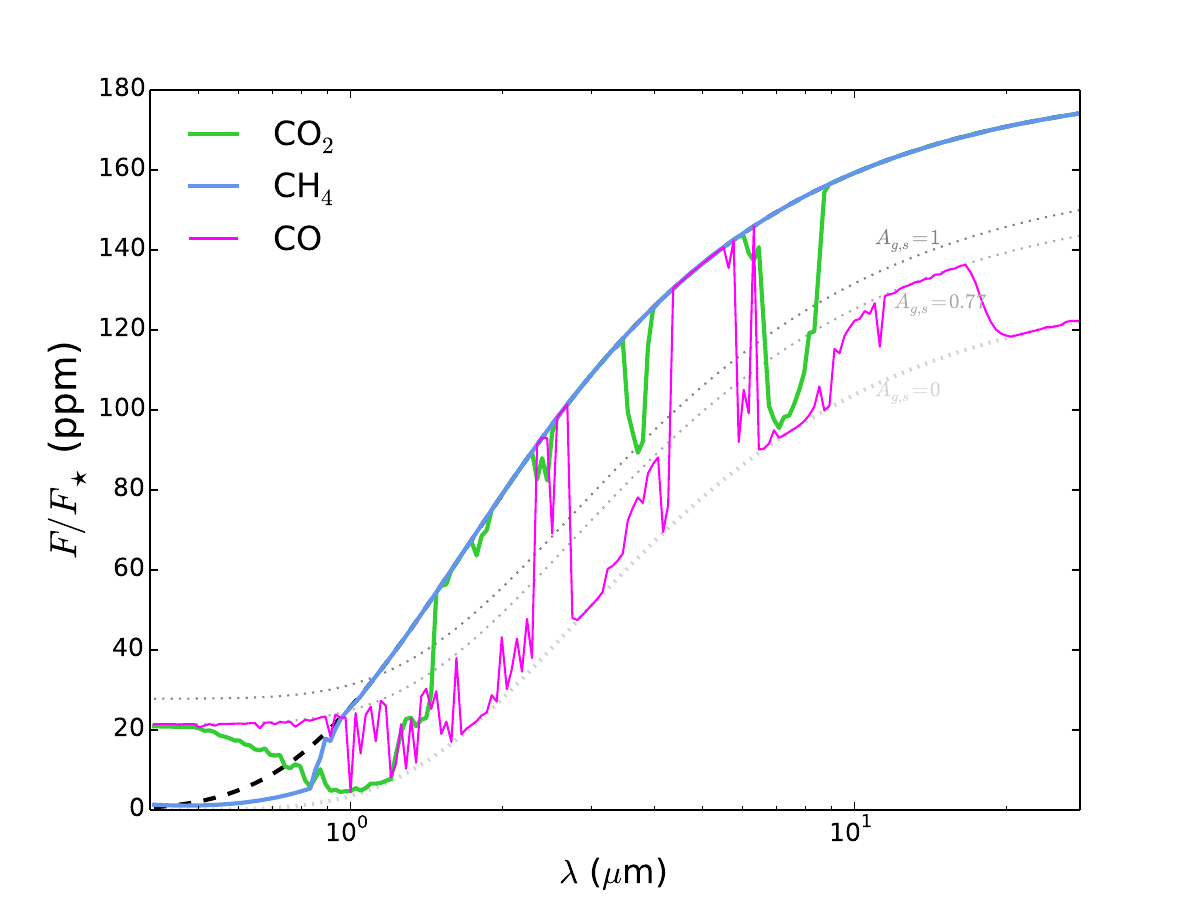}
\end{center}
\vspace{-0.1in}
\caption{Emission spectra (wavelength-dependent secondary eclipse depths) for $P=0.1$ bar, $T=2700$ K and $T_s=1965$ K.  The gray dotted curves represent ``bare rock" models with various assumed values for the geometric albedo of the rocky surface ($A_{g,s}$).  The $A_{g,s}=0$ curve essentially corresponds to a 1965 K blackbody.  For Rayleigh scattering, $A_g=0.77$ when $\omega=1$.  The black dashed curve corresponds to a 2700 K blackbody.}
\label{fig:eclipse}
\end{figure}

\begin{figure}
\vspace{-0.1in}
\begin{center}
\includegraphics[width=\columnwidth]{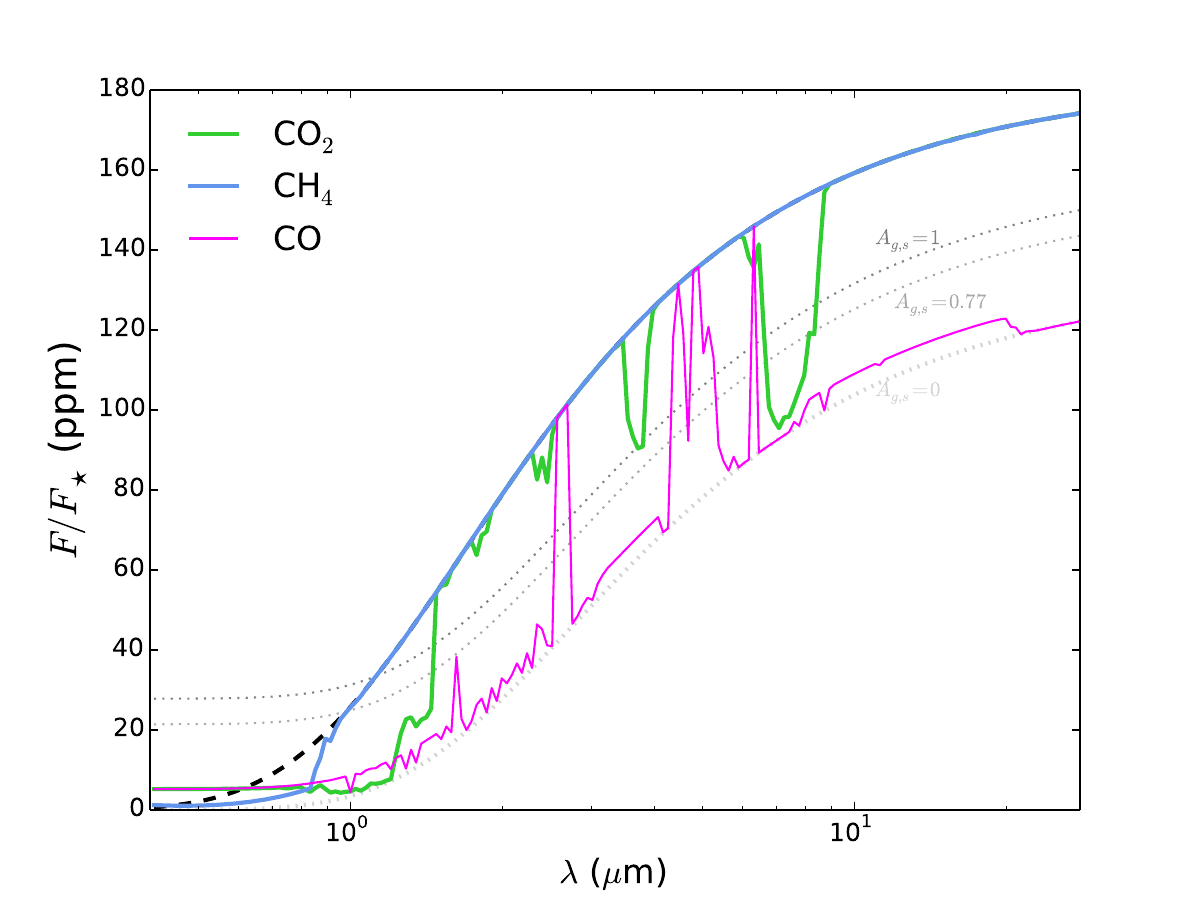}
\end{center}
\vspace{-0.1in}
\caption{Same as Figure \ref{fig:eclipse}, but for $P=1$ mbar.  Since this atmosphere is optically thin to Rayleigh scattering, the treatment of multiple scattering is ignored; see equation (\ref{eq:Ag_single}).}
\label{fig:eclipse2}
\end{figure}

\subsection{Emission spectra}

Figure \ref{fig:eclipse} shows examples of emission spectra from 0.4 to 28 $\mu$m, which cover the range of wavelengths probed by CHEOPS, TESS, Spitzer and JWST.  For illustration, these models assume an atmospheric surface pressure of $P=0.1$ bar, an atmospheric temperature of $T=2700$ K and a surface temperature of $T_s=1965$ K.  To guide our intuition, I first compute ``bare rock" emission spectra where the rocky surface is assumed to have a constant geometric albedo $A_{g,s}$.  Even with $A_{g,s}=1$, the secondary eclipse depth is only 28 ppm, although this depends on the assumed value of $T_s$.  The $A_{g,s}=0$ curve essentially corresponds to a blackbody curve with a temperature of 1965 K.

All of the computed emission spectra have spectral features that are bracketed by the 1965 K and 2700 K blackbody curves.  At wavelengths that are opaque to radiation, only the $T=2700$ K atmosphere is visible to the observer.  At wavelengths that are transparent to radiation, the observer sees the $T_s=1965$ K rocky surface.  These qualitative insights remain even when other values are assumed for $T$ and $T_s$.

As already anticipated, methane absorbs more strongly in the infrared range of wavelengths compared to CO and CO$_2$.  Its emission spectrum is featureless because it probes the thin but opaque atmosphere.  Methane scatters weakly in the optical/visible range of wavelengths with the total eclipse depth (which includes thermal emission) being less than 5 ppm for wavelengths shorter than 0.8 $\mu$m.  

With its transparent spectral windows, carbon monoxide produces an emission spectrum rich with features that rise above the continuum by as much as 60 ppm.  In the optical/visible range of wavelengths (0.4--0.8 $\mu$m), it produces a maximum eclipse depth of 22 ppm.  

Carbon dioxide is intermediate between CO and CH$_4$, producing an emission spectrum that has a few spectral features and a maximum eclipse depth of 21 ppm from 0.4--0.8 $\mu$m.  It is noted that these optical/visible eclipse depth estimates depend on the value of $T$ assumed as hotter atmospheres contribute more thermal emission at these wavelengths and thus produce larger eclipse depths.

\subsection{Eclipse depths}

\begin{table}
\vspace{-0.1in}
\label{tab:depths}
\begin{center}
\caption{Bandpass-integrated eclipse depths (ppm)}
\begin{tabular}{lccc}
\hline
\hline
Bandpass & Pure CO$_2$ & Pure CO & Pure CH$_4$ \\
\hline
CHEOPS (blackbody) & 14.8 & 21.4 & 4.0 \\
CHEOPS (SED) & 14.8 & 21.5 & 4.1 \\
TESS (blackbody) & 11.0 & 21.3 & 7.8 \\
TESS (SED)  & 11.2 & 21.4 & 7.9 \\
Spitzer 4.5 $\mu$m (blackbody) & 101.5 & 88.6 & 101.4 \\
Spitzer 4.5 $\mu$m (SED) & 107.0 & 93.4 & 107.0 \\
\hline
\hline
\end{tabular}
\end{center}
Note: All of these estimates assume a surface pressure of $P=0.1$ bar.  ``SED" means a PHOENIX spectrum of 55 Cancri was used to compute the eclipse depth.
\end{table}

To facilitate comparison to observations, the bandpass-integrated eclipse depth $D$ needs to be computed for each model emission spectrum.  Some attention to detail is needed when integrating equation (\ref{eq:master}) over wavelength as one needs to include the filter response function $f_\lambda$ of the various space telescopes\footnote{\texttt{http://svo2.cab.inta-csic.es/theory/fps/}}.  The thermal emission component of the eclipse depth is
\begin{equation}
D_{\rm th} = \left( \frac{R}{R_\star} \right)^2 \frac{\int \left[ B_\lambda \left(T_s\right) {\cal T} + \left( 1 - {\cal T} \right) B_\lambda\left(T\right) \right] S_\lambda ~d\lambda}{\int B_\lambda \left(T_\star \right) S_\lambda ~d\lambda},
\end{equation}
where $S_\lambda = \lambda f_\lambda$ for photon counters (CHEOPS) and $S_\lambda = f_\lambda$ for energy counters (TESS and Spitzer).  The reflected light component of the eclipse depth requires even more attention to detail, because a bandpass-integrated geometric albedo is similar\footnote{Geometric and spherical albedos are wavelength-dependent quantities that are intrinsic to the scattering surface or atmosphere.  They do not depend on the stellar properties.  By contrast, the Bond albedo depends on both the intrinsic scattering properties \textit{and} the spectral energy distribution of the star.  In other words, the exact same exoplanet orbiting stars of different spectral types will have different values of the Bond albedo.} to a Bond albedo (e.g., \citealt{heng21}) and requires an extra weighting factor of the stellar spectral energy distribution,
\begin{equation}
D_{\rm r} = \left( \frac{a}{R} \right)^2 \frac{\int A_g B_\lambda \left(T_\star \right) S_\lambda ~d\lambda}{\int B_\lambda \left(T_\star \right) S_\lambda ~d\lambda}.
\end{equation}
The bandpass-integrated eclipse depth is $D = D_{\rm th} + D_{\rm r}$.

There is a rich debate on how the treatment of the spectral energy distribution of the 55 Cancri star affects the interpretation of the secondary eclipses of 55 Cancri e \citep{crossfield12}.  To check the sensitivity of the computed eclipse depths to this issue, the PHOENIX model spectrum of 55 Cancri used by \cite{demory23} and \cite{mv23} is used.  The preceding expressions for $D_{\rm th}$ and $D_{\rm r}$ are generalised by substituting $B_\lambda(T_\star)$ with $F_\star/\pi$, where $F_\star$ is the PHOENIX model spectrum (in flux, rather than intensity, units).

Table 1 states the computed eclipse depths in the CHEOPS, TESS and Spitzer 4.5 $\mu$m bandpasses.  Only the Spitzer eclipse depths are sensitive to whether a blackbody or PHOENIX model is used for the spectral energy distribution of the star.  The computed TESS eclipse depths are consistent with the $15 \pm 4$ ppm and $8 \pm 5$ ppm measurements reported by \cite{mv22}.  To within two standard deviations, the computed Spitzer eclipse depths are consistent with the $154 \pm 23$ ppm measurement of \cite{demory16b} and the non-zero values reported in Table 4 of \cite{tamburo18}.  Also to within two standard deviations, the computed CHEOPS eclipse depths for pure CO or CO$_2$ atmospheres are consistent with almost all of the non-zero values shown in Figure 3 of \cite{mv23}.  A more definitive comparison requires JWST spectra to break the degeneracy in assumed composition that Spitzer data alone cannot provide.

\subsection{Change in infrared transit depth}

The pressure corresponding to the wavelength-dependent transit chord is \citep{hk17}
\begin{equation}
\begin{split}
P_{\rm transit} &\sim P_{\rm photosphere} \sqrt{\frac{H}{R}} \\
&\approx 0.055 P_{\rm photosphere} \left(\frac{T}{2700 \mbox{ K}}\right)^{1/2} \left( \frac{m}{28 \mbox{ amu}}\right)^{-1/2},
\end{split}
\end{equation}
where $H$ is the (isothermal) pressure scale height and the gravity and exoplanetary radius are taken from \cite{demory16b}.  The preceding estimate focuses on CO, because it has a larger pressure scale height compared to that of CO$_2$ and the photospheric pressure is as low as $P_{\rm photosphere} \sim 0.1 ~\mu$bar (bottom right panel of Figure \ref{fig:quantities}).  The pressure associated with the transit chord thus reaches as high in altitude as $\sim 1$ nbar.  For a $P = 0.1$ bar atmosphere, the difference between the transit radius and the surface of the exoplanet is about $\delta \approx 18 H$.  This corresponds to a \textit{maximum} change in transit depth of $2 R \delta/R^2_\star \approx 38$ ppm, which is not inconsistent with the variation in the Spitzer 4.5 $\mu$m transit depths reported by \cite{demory16a} and \cite{tamburo18}.

\section{Discussion}
\label{sect:discussion}

I now suggest an alternative interpretation of the observations reported by \cite{demory16a,demory16b,demory23}, \cite{tamburo18} and \cite{mv23}.  When the optical/visible secondary eclipse depths are consistent with being zero, this is interpreted as the observations probing mainly the bare rocky surface of the dayside of 55 Cancri e.  To be consistent with zero optical/visible eclipse depth, the geometric albedo of the surface must be close to zero.

Stochasticity in the outgassing and atmospheric escape fluxes plausibly lead to fluctuations in the global spatial distribution of the atmosphere (which I have not modelled) and the atmospheric temperature.  As the outgassed atmosphere starts to accumulate on the dayside, this increases the optical/visible eclipse depth because of Rayleigh scattering of starlight \textit{and} atmospheric thermal emission.  In this scenario, pure CH$_4$ atmospheres are ruled out because they produce insufficient Rayleigh scattering, although such atmospheres would easily produce a featureless blackbody spectrum in the infrared.  CO and CO$_2$ atmospheres are also opaque at 4.5 $\mu$m and the emission spectrum at this wavelength probes the $T=2700$ K blackbody curve.

The fluctuating temperatures shift the upper envelope of the infrared emission spectrum, shown in Figure \ref{fig:eclipse}, up and down.  This produces variable infrared eclipse depths that are consistent with those reported by \cite{demory16a} and \cite{tamburo18}.  It is plausible that the changing atmosphere also causes erratic fluctuations in the optical/visible phase curves, which have been observed by CHEOPS \citep{mv23}.  The infrared eclipse depth never becomes zero in this scenario, because even in the absence of an atmosphere it probes the rocky surface.

Upcoming JWST observations will potentially be able to distinguish between different atmospheric chemistries.  If the atmospheric surface pressure is less than 0.1 bar (see Figure \ref{fig:eclipse2} for computed emission spectra corresponding to $P=1$ mbar), then the spectral features will become more distinct.  However, the spectral features may easily be muted by the presence of clouds, hazes or condensates.  Nevertheless, fitting even a featureless spectrum to a blackbody curve will allow one to infer the atmospheric temperature, independent of the surface pressure.  If spectral features are observed, then fitting a blackbody curve to the lowest fluxes will allow one to infer the surface temperature.  A model fit should be performed within a Bayesian retrieval framework with the fitting parameters being the atmospheric temperature ($T$), surface pressure ($P$) and surface temperature ($T_s$).

The current model makes falsifiable predictions that may be tested by simultaneous optical/visible and JWST observations.  \textit{It is crucial that these observations are taken at the same time, because the atmospheric escape and radiative adjustment timescales are expected to be shorter than the orbital period.}  When 55 Cancri e is in its ``bare rock" phase, the optical/visible eclipse depth should be close to zero while the infrared emission spectrum should probe the temperature of the rocky surface.  When an outgassed atmosphere is present on the dayside, the optical/visible eclipse depth probes Rayleigh scattering by the atmosphere while the infrared emission spectrum should probe the atmospheric temperature and composition.  In between these two phases, the infrared and optical/visible eclipse depths are expected to fluctuate as the atmospheric temperature adjusts to radiative equilibrium (from its original outgassed value).

It is conceivable that the molecules of the outgassed atmosphere will eventually be broken up into their constituent atoms and ions (e.g., carbon and oxygen), which may be observable as exospheric species via ultraviolet spectroscopy.

\vspace{0.2in}
{\scriptsize I am grateful to Alexis Brandeker, Brice-Olivier Demory and Meng Tian for useful discussions and to Erik Meier Vald\'{e}s for providing the PHOENIX stellar spectrum of 55 Cancri.  I acknowledge funding by the Ludwig Maximilian University and the Swiss National Science Foundation.  This research was stimulated by spirited discussions at the inaugural \textit{Exoplanets by the Lake} summer school, held from 31st July to 4th August 2023 in Starnberg, following a talk by Jayshil Patel.}

\label{lastpage}

\end{document}